\documentclass[showpacs,pre,aps]{revtex4}
\usepackage{graphicx}
\def\be{\begin{equation}}
\def\ee{\end{equation}}
\def\bea{\begin{eqnarray}}
\def\eea{\end{eqnarray}}

\begin{document}
\title{Riemann zeros, prime numbers and fractal potentials}
\author{Brandon P. van Zyl}
\affiliation{Department of Physics and Astronomy,
McMaster University, Hamilton,
Ontario, Canada, L8S~4M1}
\author{David A. W. Hutchinson}
\affiliation{Department of Physics, University of Otago, P.O. Box 56,
Dunedin, New Zealand}

\begin{abstract}
Using two distinct inversion techniques, the local one-dimensional 
potentials for the Riemann
zeros and prime number sequence are reconstructed.  
We establish that 
both inversion techniques, when applied to the same set of levels,
lead to the same fractal potential.  This provides numerical
evidence that the potential
obtained by inversion of a set of energy levels is unique in one-dimension.
We also investigate the fractal properties of the
reconstructed potentials and estimate the fractal dimensions to be  
$D=1.5$ for the Riemann zeros and $D = 1.8$ for the prime numbers.
This result is somewhat surprising since the nearest-neighbour spacings
of the
Riemann zeros are known to be chaotically distributed whereas the primes
obey almost poisson-like statistics.  
Our findings show that the fractal dimension is
dependent on both the level-statistics and spectral rigidity, $\Delta_3$,
of the energy levels.
\end{abstract}

\maketitle
\section{Introduction}
\label{intro}

In the early work of Wu {\em et. al} \cite{wu1} a 
general numerical algorithm was developed which
could reconstruct a one-dimensional (1D) local potential from an essentially
arbitrary set of levels (i.e., bound-states).  
The basic idea behind the method (see Sec.~II below) is to
functionally minimize the potential so that it reproduces the
prescribed energy eigenvalues to numerical accuracy.
This is, in principle, a solution of the 1D quantum 
inverse problem for a confined system if one could fit
the infinitely many levels to a local potential.  Of course, this is 
beyond any numerical algorithm's ability, and one has to settle for
fitting only the first $N$ eigenvalues.  Since the algorithm requires
a direct solution of the Schr\"odinger equation (SE) for every eigenvalue, 
it is practically limited to fitting roughly $N \sim {\cal O} (10^3)$ 
eigenvalues.
In an interesting application of this algorithm, Wu 
and Sprung \cite{wu2} chose as their eigenvalue
spectrum the complex zeros (it is implicit throughout this paper that zeros
refer to the complex zeros of the zeta function) 
of the Riemann zeta function, which is well known
in number theory \cite{edwards}.  By finding a local 1D potential which 
exactly reproduced the first $N = 500$ ``eigenvalues'', they
had effectively found a 1D quantum Hamiltonian for the first 500
zeros of the
Riemann zeta function.  This is of course interesting in its own right
because such Hamiltonians have long been conjectured to
hold the key to proving the celebrated Riemann hypothesis
\cite{edwards}.
Their reconstructed potential was found to be fractal, and the fractal
dimension was estimated to be $D = 1.5$.   
This result suggests
the intriguing possibility that an integrable system 
with time-reversal symmetry may generate the Riemann zeros provided the
potential is fractal.  It is worth commenting that the precise meaning one
can attach to such 1D potentials is not clear.  In particular,
taking a chaotic system (i.e., the Riemann zeros) 
and forcing it to be integrable
will lead to a potential that is dependent on the dimension of the Hilbert
space (i.e., number of levels fitted).

Motivated by the work mentioned above, Ramani {\em et.~al}\cite{ramani1}
also investigated the reconstruction of a 1D system's quantum Hamiltonian from
its eigenvalues.  Their approach to solving the inverse problem was to use 
a ``dressing transformation''
based on techniques developed in the solution of the nonlinear 
Korteweg-de Vries equation.
Their method, being much more efficient numerically than the direct SE 
approach resulted in many more levels being fitted for the 
same numerical effort.
Unfortunately, rather than directly examining the Riemann zeros, as Wu 
and Sprung had previously done, 
they investigated the fractal properties of potentials generated from energy
levels whose nearest-neighbour spacing distribution (NNSD) were the
Gaussian Unitary Ensemble (GUE) \cite{stockmann}.   Although 
the NNSD of the Riemann zeros are also known to belong to the GUE,
Ramani {\em et. al} could only make general comments 
about 1D potentials generated from the GUE statistics, but nothing specific about
the Riemann zeros.  Nevertheless, their work suggests, 
using ${\cal O}(10^4)$ levels, a fractal dimension of $D=2$ as
$N \rightarrow \infty$ for GUE
reconstructed potentials. By extension, they conjectured that
their more accurate estimate of the fractal dimension 
leads to a value of $D=2$ for the Riemann potential.

In a reply to this work, Wu and Sprung \cite{wu3}
have pointed out that using a
different spectrum (i.e., a set of energy levels whose NNSD is the GUE 
rather than the actual complex Riemann zeros) does not 
necessarily imply that the fractal dimension 
of the reconstructed potential should
be the same.  One of the key reasons behind this statement is the
lack of long-range correlations (i.e., the so-called spectral rigidity 
$\Delta_3$) in the 
generic GUE spectrum used in Ref.~\cite{ramani1}.  
Indeed, it was argued that
if the fitting spectrum does not contain proper long-range correlations,
the resulting potential may appear to be more random than
the Riemann zeros potential.  
This, of course, can result in two different estimates for the fractal
dimension even if a large number of levels have been fitted.
Another important concern raised in their reply 
is that the potential
obtained by inversion of the levels need not be unique.  In other words,
even if Ramani {\em et. al} had used the Riemann zeros as their fitting 
spectrum, 
they may not have obtained the same potential; that is,
the fractal dimension may depend on the inversion technique applied.

In light of the above, we believe that there are several 
questions which have been
left unanswered: (i)  For the same
set of energy levels, are the potentials reconstructed from different
inversion procedures unique ? 
(ii) Does the fractal dimension depend on the inversion technique ?
(iii) What is the fractal dimension for the Riemann zeros potential ?
(iv) Is the fractal dimension a generic quantity for
reconstructed 1D potentials or does it depend on the NNSD and $\Delta_3$
statistics of the energy level spectrum ?

In what follows, we will attempt to provide 
answers to these questions. 
We will begin by presenting a brief overview of the numerical
inversion techniques used in Refs.~\cite{wu2,ramani1}, 
and then apply them to reconstruct the potentials of the Riemann zeros
and prime numbers.  
Our decision to focus on these number sequences
is three-fold.  First, the contradictory results $D=1.5$ and $D=2$
for the fractal dimension of
the Riemann potential remains unresolved.  
Secondly, the Riemann zeros are known to 
be dual to the prime numbers in the sense that all of the primes are
encoded through the complex Riemann zeros via Riemann's formula 
\cite{edwards,bhaduri}.
One may then speculate as to whether any additional relationships
between the Riemann zeros and prime numbers can be obtained 
from the reconstructed
fractal potentials.  Finally, since the NNSD of the Riemann zeros 
belong to the GUE class,
and the prime numbers obey a more poisson-like distribution, it is 
interesting to investigate the dependence of the fractal dimension
on the level statistics.
The reconstructed potentials obtained from
the two inversion methods will then be used to address the questions of
uniqueness and fractal dimension.
\section{Reconstructing the potential from a set of levels}
In this section, we will give a brief overview of the two inversion techniques
used in this paper to reconstruct the local 1D potentials.   
After this presentation, we will compare
the reconstructed potentials for the Riemann zeros and prime number
sequence obtained from both methods.
\subsubsection{Direct SE approach}
Perhaps the most intuitive approach to solving the inverse problem is
to directly invoke the SE.  One initially starts from some arbitrary potential
$V(x)$
(although for reasons of convergence, it is wiser to start from the 
semiclassical potential as described in Ref.~\cite{wu2}) 
and solve the SE to obtain a set of eigenvalues
$\varepsilon_n$.  One then defines a ``cost function''
\be
F = \sum_n(\varepsilon_n - E_n)^2~,
\ee
where the $E_n$ are the exact eigenvalues one wishes to reproduce.
The potential $V(x)$ is now adjusted in such a way as to minimize the
cost function.  Mathematically, this results in the following functional
equation:
\be
\frac{\delta F}{\delta V(x)} = 2\sum_n (\varepsilon_n - E_n)\phi^2_n(x) = 0~,
\ee
where $\phi_n(x)$ is the $n$-th normalized wavefunction.
The functional minimization can be performed using a conjugate gradient
method and the Numerov technique can be used to solve the discrete 1D SE.  
We have used this approach to successfully reproduce all of the 
numerical data for the Riemann zeros reported in  Ref.~\cite{wu2}.  
Unfortunately, even on modern 
PC/Workstations, this direct approach involves a large computational
undertaking.  We have found it impractical to go beyond 
$N=2000$ levels when using this technique.
\subsubsection{Method of the dressing transformation}
The basis for the dressing transformation can be found in the soliton
theory of nonlinear wave equations.  The details of this approach are
presented in Ref.~\cite{ramani1}, 
and we give here only the essential equations required
for its implementation.  The starting point is a given set of
levels $\epsilon_0, \epsilon_1,...,\epsilon_{N-1},\epsilon_{N}$ which have been shifted
so that one has exactly $N-1$ negative eigenvalues with the
last eigenvalue satisfying $\epsilon_N = 0$.  In this way, the potential
is constructed from the ``top down''.  This does not impose any restrictions
other than the requirement that the number of levels to be fitted is fixed
in advance.  An initial potential $V(x)=0$ is then used as input and the
first-order differential equation (diffeq)
\be
f' - f^2 + V(x) = \epsilon
\label{feq}
\ee
is evaluated numerically with $f(0) = 0$ and $\epsilon = \epsilon_{N-1}$.
Once the function $f$ has been found at all points, a new potential
$W(x)$ is constructed from the diffeq:
\be
W(x) = 2\epsilon + 2f^2 - V(x)~.
\ee
This new potential is then substituted back into Eq.~(\ref{feq}) 
(i.e., $W(x) \rightarrow V(x)$) with the eigenvalue now being
$\epsilon = \epsilon_{N-2}$.  
The choice $f(0)=0$ is enforced throughout the calculation and ensures that
the final potential $V(x)$ is even.
This iterative procedure is continued until all of the eigenvalues have
been exhausted.  The final $W(x)$ is the desired reconstructed potential
which reproduces exactly all of the $N$ eigenvalues.  The integration of
Eq.~(\ref{feq}) can be performed using any standard method, but we have
used a fourth-order Runge-Kutta (RK) method with step
size $h = 1\times 10^{-5}$ to ensure high accuracy.  
Nevertheless, it should be clear that the dressing transformation is numerically
a far superior method since it does not require a large parameter search
in the construction of the potential.  Consequently, it is easy to fit
orders of magnitude more levels than in the direct SE approach.
\begin{figure}
\rotatebox{-90}{
\resizebox{4in}{5in}{\includegraphics{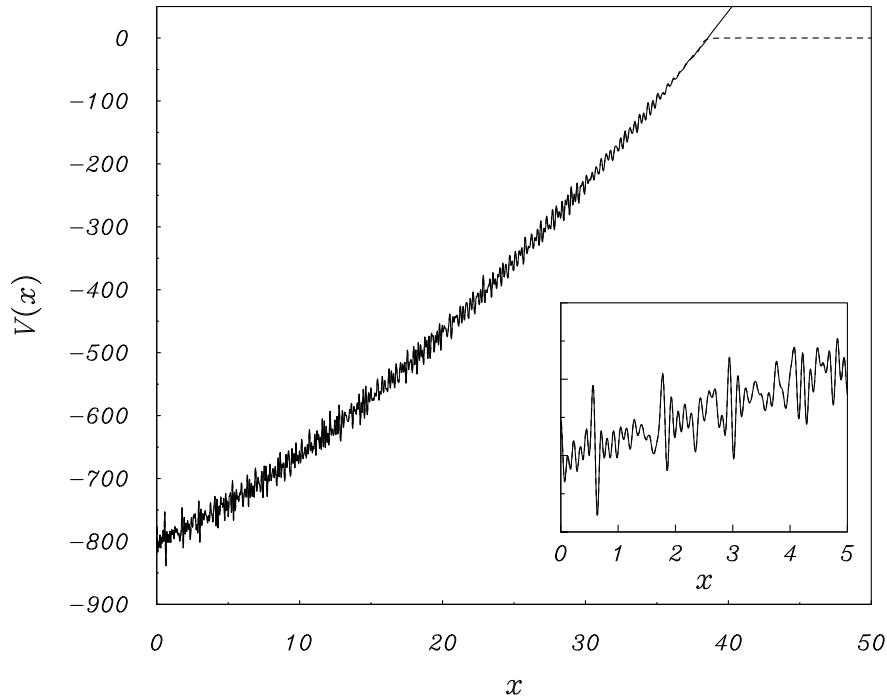}}}
\caption{The reconstructed potential for the Riemann zeros with $N=500$
fitted levels.  The dashed curve is obtained from the dressing transformation
while the solid curve is from the direct SE approach.  The inset shows a
magnified view 
of the potential near the minimum.  The two curves are indistinguishable
below $x \approx 35$.}
\label{fig1}
\end{figure}
\begin{figure}
\rotatebox{-90}{
\resizebox{4in}{5in}{\includegraphics{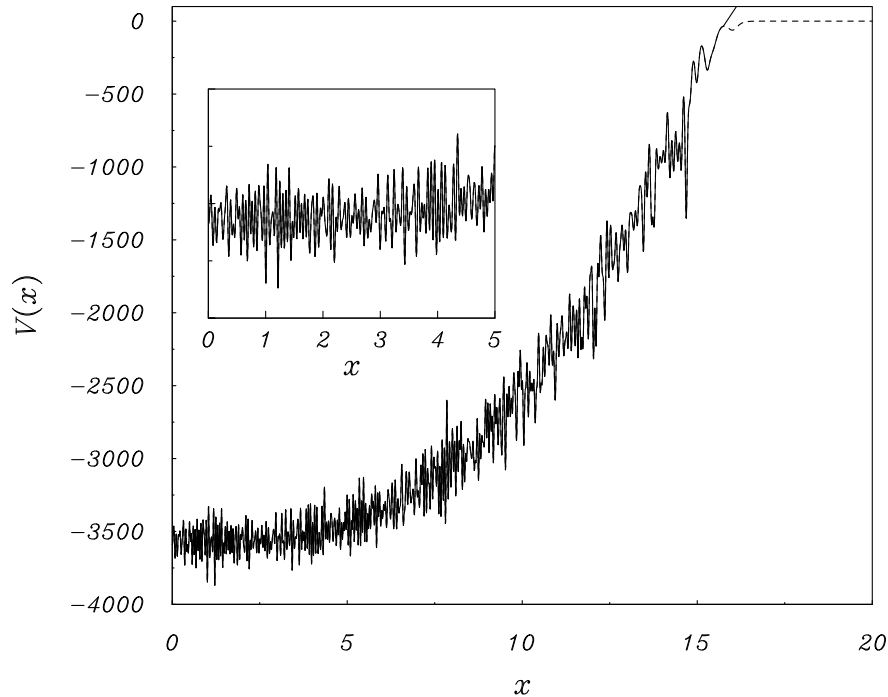}}}
\caption{As in Fig.~1, but for the prime numbers.  Again, the curves
are indistinguishable for energies below the last fitted level.}
\label{fig2}
\end{figure}
\subsubsection{Uniqueness of the reconstructed potentials}
Having provided the basic background for numerically solving the inverse 
problem,
we are now in a position to address the question of uniqueness.
To this end, we first reconstruct the potential for the Riemann zeros
with $N=500$.  This number was chosen because it represents the maximum
number of Riemann zeros fitted in Ref.~\cite{wu2}.
Figure 1 displays the
potential obtained from the direct SE approach (solid curve)
and the dressing transformation (dashed).  
In the inset to the figure, we zoom-in on the
the region $x \in [0,5]$ to illustrate the remarkable agreement between the
two methods, which
continues until one approaches energies near the last fitted level
$(x \approx 35)$.
Obviously above this threshold the potentials will not agree since the
SE approach uses the semiclassical
profile as its zeroth-order potential
whereas the dressing transformation takes
$V(x) = 0$.  However, as the number of levels is increased, this threshold 
is moved to higher and higher energies, so that as $N \rightarrow \infty$,
the two potentials should agree over all space.  
The equivalence between the
two inversion techniques has been checked for up to $N=2000$ (which
computationally speaking, represents the upper limit of the SE approach), but
we see absolutely no reason why this equivalence should not be true for 
arbitrarily large $N$.
Nevertheless, to ensure that this agreement is not in some way fortuitous, 
we have also
reconstructed the potential for the prime number sequence with $N=500$.
The results of the two approaches for the primes are shown in 
Fig.~2.  Again, the potentials are indistinguishable except for regions
near the last fitted level.  
\begin{figure}
\rotatebox{0}{
\resizebox{4in}{5in}{\includegraphics{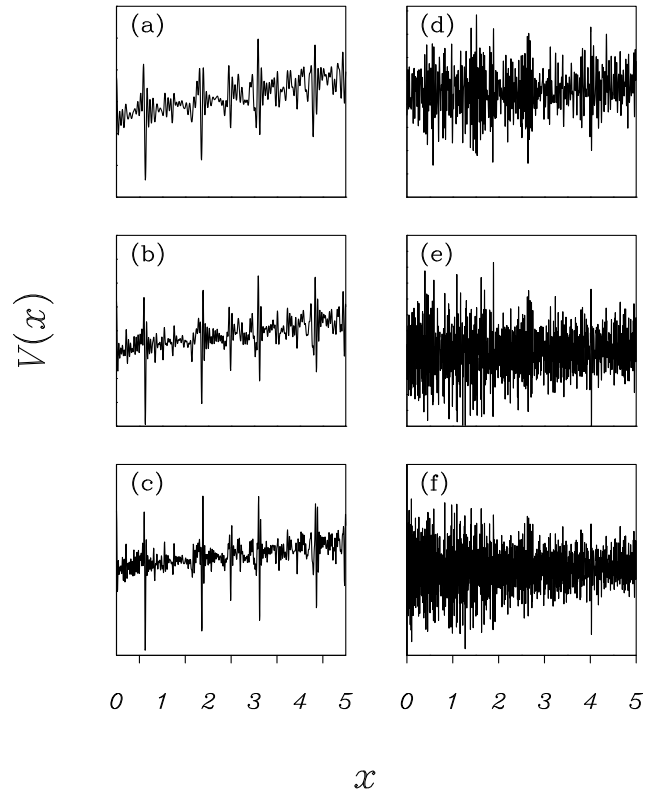}}}
\caption{Panels (a)-(c) and (d)-(f) correspond to the Riemann zeros and
prime potential, respectively.  The number of fitted levels from top to bottom
is $N=2000,~5000,~10000$.  We have zoomed in on the range $x\in [0,5]$ to
highlight the differences between the two potentials.}
\label{fig3}
\end{figure}

It also proves instructive to compare in more detail the
reconstructed potentials for the Riemann zeros and prime numbers.
To facilitate this comparison, we present in Fig. 3 details of the
potentials for $N=2000,~5000,~10000$ levels around the minimum
(one could look at any other region on the curves 
and obtain similar graphs).  Figure 3 (a)-(c)
corresponds to the Riemann zeros
whereas Fig.~3 (d)-(f) corresponds to the prime potential.  
Several interesting 
observations can
be made at this point.  First, we note that for a given number of
fitted levels, the Riemann zeros potential looks less ``noisy''
than the prime potential.  In fact, by $N=10000$, the prime potential takes
on the characteristics of (roughly speaking)
white noise whereas the Riemann zero potential appears to have far 
more structure.  In particular, the Riemann potential contains large
jumps at very specific spatial positions which appear to be quite robust
(i.e., the positions of the jumps do not
change as $N$ increases).  In contrast, any local structure
present in the prime potential is completely washed out by
$N=10000$.  This 
suggests that the prime potential may have a larger fractal dimension
(i.e., close to $D=2$ of white-noise)
than the Riemann zeros.  Indeed, these observations 
lead us to consider more closely the fractal potentials 
generated by Ramani {\em et. al} \cite{ramani1}. 

As we mentioned earlier, the potentials in Ref.~\cite{ramani1} were obtained
from a generic GUE distribution without 
accounting for the $\Delta_3$ statistics.
A comparison of our Fig.~3 (a)-(c)
with their Fig.~1 (a)-(c)
clearly illustrates that a generic GUE distribution {\em cannot}
capture, even qualitatively, the details of the Riemann zeros potential.
We are therefore inclined to agree with the assertion in Ref.~\cite{wu3}
that long-range correlations in 
the level spectrum play an important role in determining the fractal
properties of the potential. 
This is further emphasized by noting that for large $N$, the prime 
potential [see Fig.~3 (f)] looks very similar to
the large $N$ GUE fractal potential in Ref.~\cite{ramani1}, 
which was conjectured to have 
$D = 2$ as $N \rightarrow \infty$.  It is perhaps then
not so ambitious to suggest that the prime potential also has fractal dimension
$D=2$.  We cannot however make such claims for the Riemann zeros, which
evidently require a more careful quantitative analysis.
\section{Fractal dimension}
\begin{figure}
\rotatebox{-90}{
\resizebox{4in}{5in}{\includegraphics{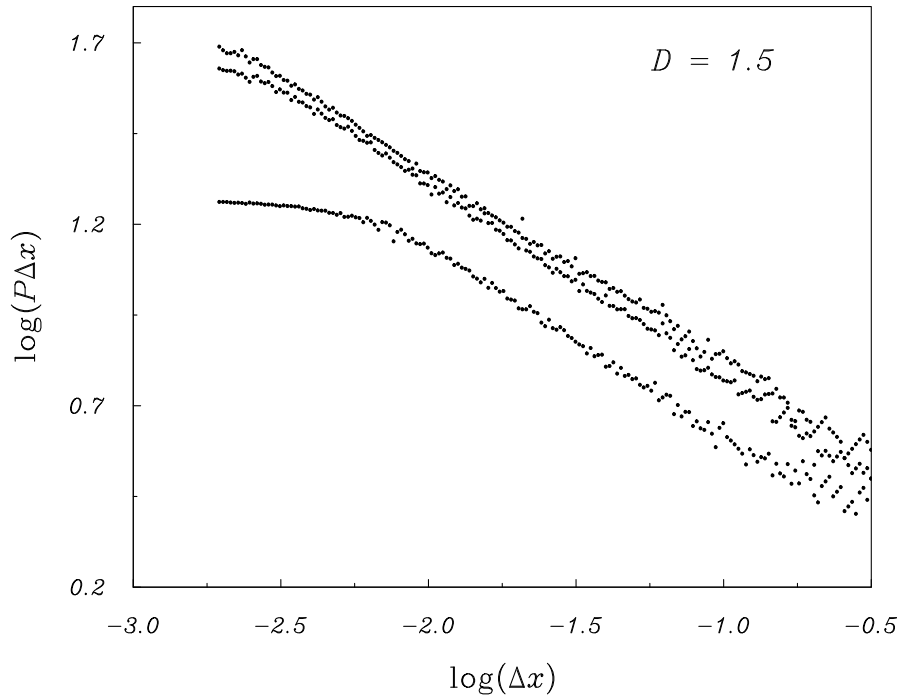}}}
\caption{Box-counting method analysis of the fractal dimension of the
Riemann potential for (top to bottom) $N=10000,~5000,~500$.  The negative 
slope of the curve is the fractal dimension minus one.} 
\label{fig4}
\end{figure}
We have seen in
the previous section that for the same number of fitted levels, the
oscillatory behaviour of the Riemann zeros and prime potentials are quite
different.  What significance then (if any) do these rapid oscillations have
on the fractal dimension ?  Before we answer that question, it is perhaps
useful to give a short account of the method used by the authors in
Refs.~\cite{wu2,ramani1} 
to estimate the fractal dimension of their reconstructed potentials.

The standard procedure for numerically estimating the fractal dimension of
curves such as the ones we have generated is to use the box-counting technique
\cite{feder}.
Simply put, the box-counting technique involves choosing a region on the
curve (over which the character of the oscillations does not change) 
and normalizing
the axes so that the region is a square of side length equal to unity.
The square region is then divided into $n^2$ cells of side length 
$\Delta x = 1/n$
and the number of cells $P$ that contain a portion of the curve is counted.
A plot of $\log(P\Delta x)$ versus. $\log(\Delta x)$ is then constructed and 
the region over which the resulting graph has a
constant slope $1-D$ determines the
box-dimension $D$; the box-dimension is usually simply
referred to as {\em the} fractal dimension.  Although it is 
not the only method for estimating the fractal
dimension of a curve, we will make use of the box-counting technique 
to allow for a sensible comparison with the results in Refs.~\cite{wu2,ramani1}

In Fig.~4, we display the results of the box-counting method as applied
to the Riemann potential for $N=500,~5000~,~10000$.  In keeping with 
Ref.~\cite{wu2}, we have focused on the range $x\in [0,10]$ for the analysis.
What is clear from the
figure is that the slope, $m$, of the linear region in the graph
is essentially unchanged for $N > 500$.  In fact, the slopes for
$N=5000$ and $N=10000$ are virtually identical.  Given that the fractal 
dimension is given by $D = 1 + |m|$, we conclude that to two significant
figures, $D=1.5$ for the
the Riemann zeros potential.  This result is consistent with
the findings of Wu and Sprung \cite{wu2}
who likewise obtained $D=1.5$.  We see no evidence at all
for $D=2$ as was suggested by the work of Ramani {\em et. al} \cite{ramani1}.

Figure 5 shows the results of the box-counting method for the prime number
potential.  In contrast to the Riemann zeros, there is a noticeable increase
in the slope of the curve (i.e., larger fractal dimension) as one increases
the number of fitted levels from $N=500$ to $N=5000$.  However, for
$10000 < N < 40000$ (not shown), the slope is found to be practically 
unchanged.  We are therefore led to conclude that the fractal dimension
of the primes is $D=1.8$.   We are, however,  open to the possibility that as 
$N \rightarrow \infty$ the fractal dimension of the prime potential 
approaches $D=2$ based on the qualitative behaviour of the
potential seen in Fig.~3 (d)-(f).  
Indeed, a similar slow convergence of the fractal
dimension was also observed in 
Ref.~\cite{ramani1} for the GUE statistics.
However, numerically investigating this convergence 
would require possibly millions of levels, which is well beyond the 
capabilities of our current computational facilities.
\begin{figure}
\rotatebox{-90}{
\resizebox{4in}{5in}{\includegraphics{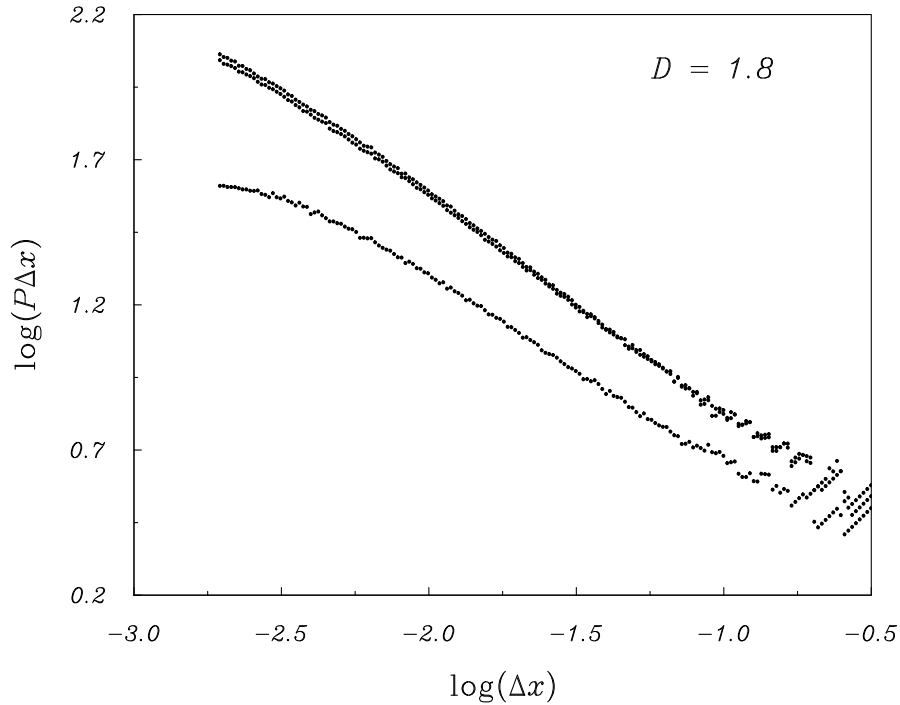}}}
\caption{Box-counting method analysis of the fractal dimension of the
prime potential with (top to bottom) $N=10000,~5000,~500$.  The negative 
slope of the curve is the fractal dimension minus one.  Note that the
difference in the curves from $N=5000$ to $N=10000$ is negligible.} 
\label{fig5}
\end{figure}
\section{Summary and Discussion}
We have numerically examined the reconstruction of a 1D local potential 
from a set of energy levels using two distinct inversion procedures.
Applying these methods to the Riemann zeros and the prime number sequence
we find that both inversion techniques produce the same quantum potential  
when applied to the same set of energy levels.
This provides
evidence that the inversion problem in 1D results in a unique potential.
Thus, the answer to the first question raised in the introduction is
affirmative.  Based on this finding, it follows that the fractal dimension
cannot depend on the method of inversion.
We have also examined the issue of the fractal dimension of the Riemann
zeros for up to $N=40000$ levels and estimate that $D=1.5$.  This
result is in complete agreement with Wu and Sprung's \cite{wu2}
earlier estimate
(albeit using only $N=500$ levels), and illustrates that the potentials
investigated by Ramani {\em et. al} \cite{ramani1}
have little bearing on the Riemann zeros potential.  
In particular, we demonstrated that it is insufficient to
simply account for the NNSD of the energy levels to capture the local
details responsible for the fractal properties of the reconstructed potential;
the Riemann zeros display local structure that is not reproduced by a
generic GUE level spectrum.
Therefore, long-range correlations appear to play a pivotal role
in determining the fractal dimension of the potential.

The prime number potential has also been investigated, and found to have
fractal dimension $D=1.8$.  This result is somewhat surprising since
the NNSD of the primes is almost poisson-like, whereas the Riemann zeros
obey the (chaotic) GUE statistics.  Unlike the Riemann potential, our
calculation of the  prime potential's fractal dimension suggests
that there may be a very slow convergence (as $N \rightarrow \infty$)
of the fractal dimension to $D=2$.  This possibility has also been noted
in Ref.~\cite{ramani1} where the fractal
properties of white noise-like potentials 
corresponding to the GUE statistics were studied.
\begin{acknowledgments}
We would like to thank Rajat Bhaduri and Donald Sprung for very fruitful
discussions.
BVZ acknowledges Prof.~Dr.~R.~K.~Bhaduri for financial support through
the NSERC of Canada.
DAWH acknowledges support from the Marsden Fund of the Royal Society of
New Zealand and a University of Otago Research Grant.
\end{acknowledgments} 
\bibliographystyle{pra}

\end{document}